# Insights into hyperbolic phonon polaritons in hBN using Raman scattering from encapsulated transition metal dichalcogenide layers

Jacob J.S. Viner [1], Liam P. McDonnell [1], Pasqual Rivera [2], Xiaodong Xu [2], David C. Smith [1*]

[1] School of Physics and Astronomy, University of Southampton, Southampton SO17 1BJ, United Kingdom

[2] Department of Physics, University of Washington, Seattle, Washington 98195, United States of America

**Abstract**
New techniques for probing hyperbolic phonon polaritons (HPP) in 2D materials will support the development of the emerging technologies in this field. Previous reports have shown that it is possible for WSe$_2$ monolayers in contact with the hexagonal boron nitride (hBN) to generate HPP in the hBN via Raman scattering. In this paper, we set out new results on HPP Raman scattering induced in hBN by WSe$_2$ and MoSe$_2$ monolayers including new resonances at which the Raman scattering is enhanced. Analysis of the observed Raman lineshapes demonstrates that Raman scattering allows HPP with wavevectors with magnitudes significantly in excess of 15000 cm$^{-1}$ to be probed. We present evidence that the Raman scattering can probe HPP with frequencies less than the expected lower bound on the Reststrahlen band suggesting new HPP physics still waits to be discovered.

**Intro**
The unique properties of hyperbolic phonon polaritons (HPPs)[1–4] are leading to their exploitation in a range of important applications[5,6]. For instance, in the production of a new class of mid-infrared sources[7] which, whilst thermally excited, produce radiation which is narrowband[8] and spatially coherent[9,10]. Work on related sources indicates that it should be possible to combine these properties with high modulation frequencies; up to 10 MHz[11]. Another field in which HPPs are being exploited is MIR integrated nanophotonics for applications in surface-enhanced infrared spectroscopy[12] as well as sub-diffraction imaging[6,13]. A recent highlight in this field has been the creation of reconfigurable waveguides and lenses for hyperbolic phonon polaritons[14,15] using phase change materials. Hyperbolic phonon polaritons allow subwavelength volume confinement[6,16,17] of mid-infrared radiation by as much as a factor of 86. This confinement is not only being exploited for miniaturisation but also allows the concentration of electromagnetic energy, allowing for strong coupling[18,19] and nonlinear effects[20–22]. By analogy with plasmonics, which are more lossy than HPP[23], an even wider range of HPP applications, e.g. biosensors[24] and improving signal to noise photodetectors[7,8], are likely to emerge soon.

The two main techniques used to study HPP are FTIR spectroscopy[25] and scattering-type scanning near-field optical microscopy (s-SNOM)[26–28]. FTIR is simpler however it either requires specially prepared microfabricated structures[5] or prism coupling[5,29] to access the large wavevector HPP. The former is not suitable for probing HPP in devices and the latter limits the wavevector that can be accessed. s-SNOM[26–28] allows HPPs to be imaged in real space with impressive resolution and thus access to wavevectors of the order of ~ 1000 times the free space wavevector[27,28]. However, this technique is complex and requires physical contact with the sample.

A HPP measurement method based upon the conversion of MIR to visible radiation would enable a wide range of new opportunities. The relative maturity of visible photonics means that such a method is likely to be much simpler. It should allow high-resolution (0.5 µm) imaging without the need for a near-field probe and because many of the key HPP are wide band gap materials it should allow imaging of sub-surface HPP. Due to the much shorter wavelength of visible light such a method might allow access to even larger wavevector HPPs. Actually, it has already been shown that transition metal dichalcogenide (TMD) layers allow the probing of HPPs via Raman scattering of visible radiation[30–32]. However, no one has explored what information about HPP can be obtained from the Raman spectra. In this paper we present significant new results on HPP Raman scattering in TMDs. These include new resonances at which the HPP Raman features are enhanced in $WSe_2$ monolayers and the first measurements of HPP Raman features in Mo based TMDs. Based upon these new results we discuss the mechanism for HPP Raman scattering; the fact that it can access much higher wavevector HPP than IR based techniques; and what Raman scattering of HPP might bring to the field of HPP technologies.

**Main Body**

Figure 1a) shows a Raman spectrum of the encapsulated $WSe_2$ monolayer taken with an excitation energy of 1.866 eV. The observed Raman features can be separated into one-phonon peaks ($E''_{TO}(M)$ at 196 cm$^{-1}$; $E''_{TO}(K)$ at 209 cm$^{-1}$; $E'_{TO}(M)$ at 229 cm$^{-1}$; $E'(\Gamma)/A_1'(\Gamma)$ at 250 cm$^{-1}$; $A''_2(M)/E'_{LO}(K)$ at 258 cm$^{-1}$.) and two-phonon peaks (range of shifts up to 500 cm$^{-1}$) and two features at around 730-850 cm$^{-1}$ and 1000-1080 cm$^{-1}$. Further discussion of the $WSe_2$ phonon Raman peaks and their attribution can be found in literature[31,33–36]. The 730-850 cm$^{-1}$ and 1000-1080 cm$^{-1}$ features have previously been associated with scattering of excitons in the monolayer by hBN HPP (lower shift feature) and a combination of a hBN HPP and a monolayer $A_1'$ phonon (upper shift feature). This hypothesis is strongly supported by the facts that no Raman scattering is observed at comparable Raman shifts[31,32] in samples where $WSe_2$ monolayers are not in intimate contact with hBN and the only peak observed in Raman spectra of hBN[30,32,37] is at 1380 cm$^{-1}$. The peaks at smaller Raman shifts, which only involve phonons, are characteristically narrower with widths of the order of 0.25 meV (2 cm$^{-1}$), whereas the HPP related features are significantly broader, at 10 meV (80 cm$^{-1}$). The Raman features can also be observed in Figure 1b) and c) where we present colourmaps of the resonance Raman spectra of an encapsulated $WSe_2$ monolayer. All the features show clear resonance behaviour in the colourmaps. For instance the Raman peak at 250 cm$^{-1}$, assigned to the degenerate $A_1'/E'$ phonons[35,38], has four resonances; when the incoming and scattered photons are resonant with the A1s and A2s bright excitonic states. As previously reported, the 730-850 cm$^{-1}$ features shows a clear outgoing resonance with the A1s state and incoming resonance with the A2s state[31,32]. As also previously reported, the 1000-1080 cm$^{-1}$ shows a strong resonance at ~1.865 eV which is both an outgoing resonance with the A1s and an incoming resonance with the A2s, i.e. a double resonance. In addition, the same features have at least two other resonances at higher laser energies which have not been reported before. The clearest of these is an outgoing resonance associated with the A2s exciton. There is also a weaker resonance, approximately 20 meV higher in energy corresponding to the A3s exciton. As more clearly shown in Figure 1b), the features at the outgoing resonances show a characteristic behaviour in which the higher shift scattering is resonant at higher laser energies. This is particularly visible from 1.95 to 2 eV. This is clear proof that the two broad features, with Raman shifts centred at approximately 800 and 1050 cm$^{-1}$, can be associated with a band of excitations rather than a single underlying excitation.

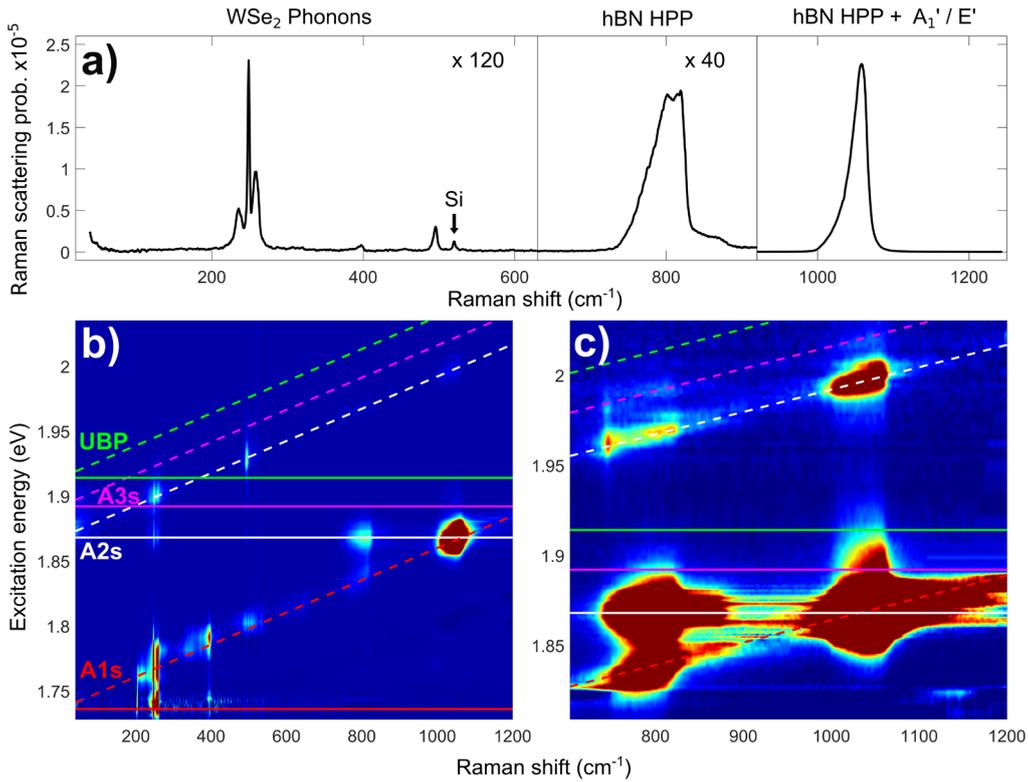

*Figure 1: a) An individual Raman spectrum from the hBN encapsulated WSe$_2$ sample with an excitation energy of 1.866 eV. The first two parts of the spectrum are scaled by a factor of 120 and 40 for visibility. b) and c) show colourmaps of the resonance behaviour of the 800 and 1050 cm$^{-1}$ hBN Raman features b) presents the full spectral width including lower Raman shift features from the WSe$_2$ phonons, such as the strong 250 cm$^{-1}$ A$_1$'(Γ)/E'(Γ) peak, as well as the higher shift hBN modes. c) Presents a zoom of the higher shift part of the spectra presented in b) with the colours adjusted to show a narrower range of intensities. The colour indicates the intensity of the Raman scattering in logarithmic scale, with blue corresponding to lowest and red to highest intensity. The energies of the incoming and outgoing Raman resonances (solid and dashed lines) associated with the A1s, A2s and A3s excitonic states are shown with red, white and pink lines respectively. The green lines correspond to the resonance conditions for the lowest energy unbound electron hole pair (UBP). The signal around 1.87 eV between the 800 and 1050 cm$^{-1}$ peaks and shifts above 1100 cm$^{-1}$ are from photoluminescence from the A1s exciton, which follows the A1s outgoing resonance.*

In order to better understand the resonance behaviour and to determine the energy of the excitonic states involved, resonance profiles of the Raman scattering were determined for the two features as presented in Figure 2. As each feature is associated with a band of excitations, the extracted resonance profile changes at different Raman shifts across the feature. For both of these hBN features, 3 resonance profiles were determined; each by integrating the Raman scattering for a 5 cm$^{-1}$ band around a centre Raman shift. This process is described in the supplementary section S1[39]. The resonance profiles presented in Figure 2 clearly show the A1s outgoing and A2s incoming resonances previously observed[30–32], as well as the new higher energy resonances. Interestingly there is no statistically significant Raman scattering associated with the hBN related features at the A1s incoming resonance. However, the significant additional noise due to the strong A1s luminescence means the upper limit on any scattering at these laser energies is comparable with the strength of the Raman scattering at the highest energy resonances. In the case of the 730-850 cm$^{-1}$ feature, shown in Figure 2a), the A1s outgoing resonance near 1.83 eV shifts to lower energy, as expected, for the lower centre Raman shifts. The two higher energy resonances above 1.95 eV show the same behaviour indicating they are also both outgoing resonances. For the 1000-1080 cm$^{-1}$ feature profiles shown in Figure 2b), the A1s outgoing and A2s incoming resonances fall at the same energy, creating a double resonance with the measured Raman scattering two orders of magnitude greater than that of the Raman feature at 730-850 cm$^{-1}$.

The resonance profiles were all fitted using the standard third-order perturbation theory prediction for Raman scattering[39] that assumes that the exciton-phonon scattering occurs in a single step. In order to fit all of the resonances, the model requires a minimum of three excitonic states. Unfortunately, this requires six exciton-phonon scattering matrix elements. This makes getting a unique fit for these parameters impossible. However, the energy and lifetime parameters for the three excitons obtained from the six independent resonance profiles, with and without setting some of the scattering matrix elements to zero, were remarkably consistent as shown in supplementary section S2[39] (see, also, references[40–42] therein). These fits give the energy (with fit estimated errors) for the three excitonic states to be 1.735 ± 0.002, 1.867 ± 0.001 & 1.892 ± 0.005 The first two energies are to within experimental error the same as the A1s and A2s energies obtained from fits to reflectivity spectra[43] and separate fits to the $A_1'$ phonon's resonance profiles shown in Figure 2c) The third energy is in agreement to within a few meV with previously published values for the A3s excitonic state[44,45] as well as values obtained from reflectivity measurements of the sample[39]. This is the first time that a Raman resonance with the A3s state has been reported.

Based upon the agreement between the measured energies of the first three excitonic energies and our model for the Rydberg series[46] we can predict the energy of the remaining excitonic states and the minimum energy unbound electron-hole states. As shown with green lines in Figure 1, the threshold for excitation of unbound electron-hole states is less than the laser energy associated with the A2s and A3s outgoing resonances. Thus, it is possible that the initial optically excited state at these resonances is a real, unbound pair excitation rather than a virtual excitonic state.

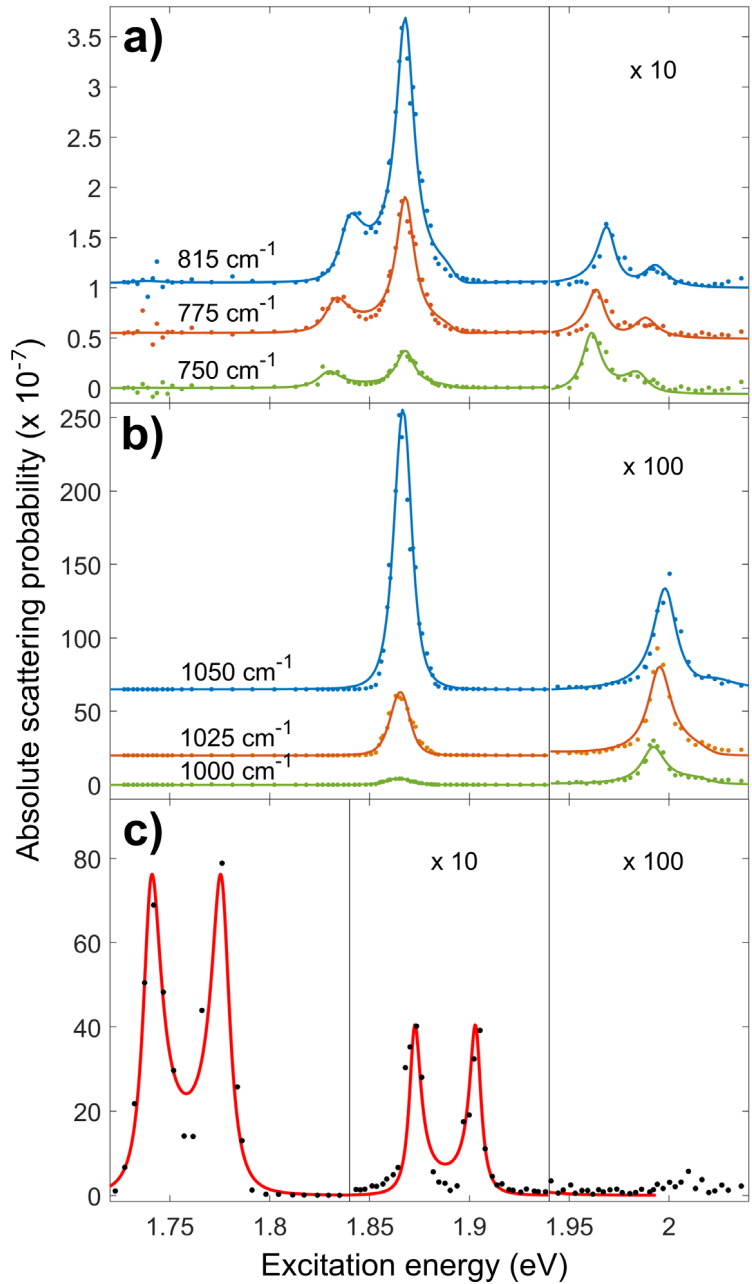

*Figure 2: Resonance profiles for the a) 800 cm$^{-1}$ or b) 1050 cm$^{-1}$ hBN Raman features determined at different Raman shifts within the feature. The insets present vertically scaled subsections of the 815 and 1050 cm$^{-1}$ resonance profiles. Peaks in the resonance profile observed in this range are associated with the A2s and A3s outgoing resonance conditions. The solid lines show fits to the resonance behaviour using 3-state, third-order, perturbation-theory resonant Raman scattering models where the three states are the A1s, A2s & A3s excitons in WSe$_2$. The resonances at ~1.84 eV and between 1.95 and 2.0 eV are outgoing resonances and thus change energy with Raman shift. The resonance at ~1.85 eV is the A2s incoming resonance. The A1s incoming resonance would fall around 1.74 eV but no statistically significant Raman signal is observed. The noise in the resonance data at the energy of the A1s exciton is photoluminescence signal from the A1s exciton and trion. c) Resonance profiles of the WSe$_2$ 250 cm$^{-1}$ single gamma point phonon over the same energy range. The left side of the plot shows the A1s incoming and outgoing resonances. The right shows the A2s resonance with intensity scaled up by a factor of 20.*

One of the key characteristic features of the hBN related Raman scattering are their broad lineshapes. However, as shown in Figure 3, the observed lineshape depends on the resonance and the energy of the laser relative to the peak of the resonance. This is more obvious at the outgoing resonance but also true at the incoming resonances. The qualitative differences between the resonances can be rationalised as being due to the variation of the resonance conditions across the lineshape and how these depend on laser energy. For instance, the relative suppression of the higher Raman shift scattering at an outgoing resonance for lower laser energies is because the lower shift scattering is closer to resonance with the outgoing state. However, attempts to quantitatively correct for this effect to obtain a single underlying lineshape did not produce a consistent lineshape independent of the laser energy. Despite this, it is possible to draw some general conclusions about the underlying lineshape. In particular, for the 730-850 cm$^{-1}$ feature the lineshape covers the whole of the Reststrahlen band. In fact, at the low shift side of peak the lineshape continues smoothly beyond the Reststrahlen band. In general, the highest scattering is closer to the upper bound of the Reststrahlen band however the exact form of the peak varies with the resonance. The lineshape of the 1000-1080 cm$^{-1}$ feature is very similar to the 730-850 cm$^{-1}$ feature. In this case spanning, and at lower shifts extending beyond, the Reststrahlen band shifted up by 250 cm$^{-1}$ as would be expected if this feature is a due to an emission of a combination of the HPP responsible for the 730-850 cm$^{-1}$ feature and the $A_1'$/E' phonon.

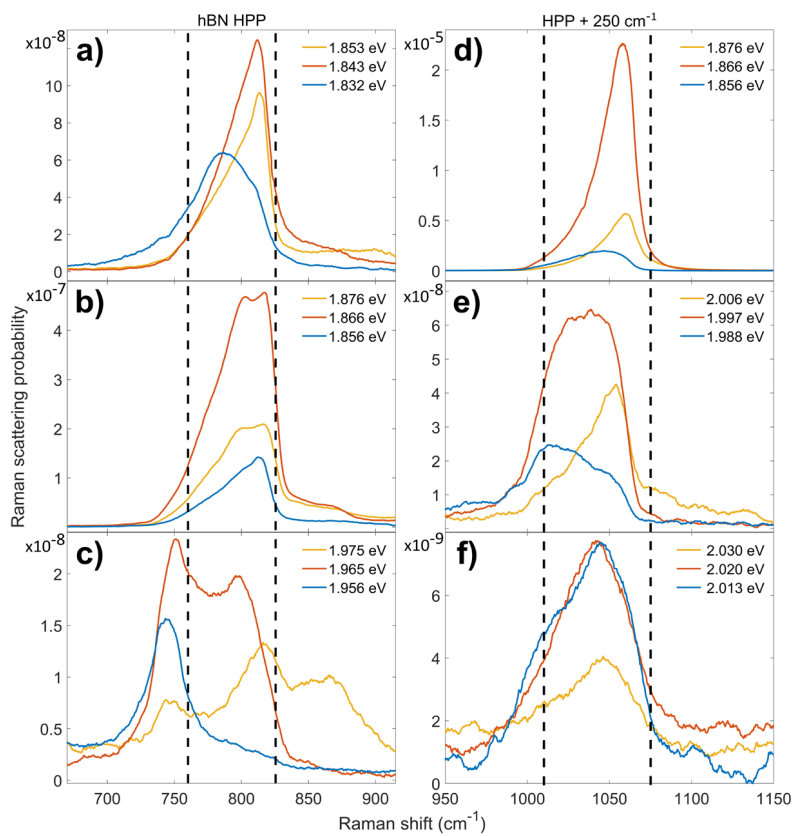

*Figure 3: Raman spectra showing the lineshapes of the 800 [a, b, c)] and 1050 cm$^{-1}$ [d, e, f)] hBN associated Raman peaks at various resonances. The orange curve shows the spectrum at the peak of the specified resonance and the blue and yellow curves show spectra taken at ± 10 meV either side of the resonance. a) A1s outgoing b) A2s incoming c) A2s outgoing d) A1s outgoing = A2s incoming e) A2s outgoing f) A3s outgoing. The lower shift peak observed in c), in the 1.956 eV spectra, at about 750 cm$^{-1}$ is associated with a multi-phonon peak rather than hBN HPP as can be seen more clearly in Fig 1c. This is only observed at the A2s outgoing resonance. The agreement of the shift of this peak with three times the shift of the $A_1'$ peak suggests that it is due to emission of 3 $A_1'$ phonons.*

Having fully analysed the data for the encapsulated $WSe_2$ we now turn our attention to the results for encapsulated $MoSe_2$. The strength of the hBN related Raman peaks in encapsulated $MoSe_2$ is significantly weaker; Fig 4. In $MoSe_2$, the features are observed only when the laser energy is tuned to the A2s or B2s excitons. In both cases, the resonances are near double resonances with the A1s and B1s excitons respectively however not fully doubly resonant. The fact that in $MoSe_2$ the resonances are significantly further from double resonance compared to $WSe_2$ is likely to be the main reason why these features are so much weaker in $MoSe_2$. Due to the weak scattering, it was not possible to obtain clear spectra apart from at the peak of the resonance or fully analyse the resonance behaviour. However, there is considerable new information available from the resonant spectra. In particular, the upper shift feature in $MoSe_2$ is considerably broader than the lower shift feature. As shown on Fig 4, it is possible to explain this if the upper shift feature is in fact due to a combination of a HPP and either an $A_1'$ (lower Raman shifts) or E' phonon (higher Raman shifts). In the case of the $WSe_2$ these phonons are degenerate and so it is possible that the upper shift feature also involves a contribution from the E' phonon. In addition, and importantly, the lineshape of the lower shift feature is very similar to the same feature in $WSe_2$ and in particular extends beyond the Reststrahlen band at lower shifts. Thus, the observation of Raman scattering outside of the Reststrahlen band is not a feature of a single material or sample but is more general.

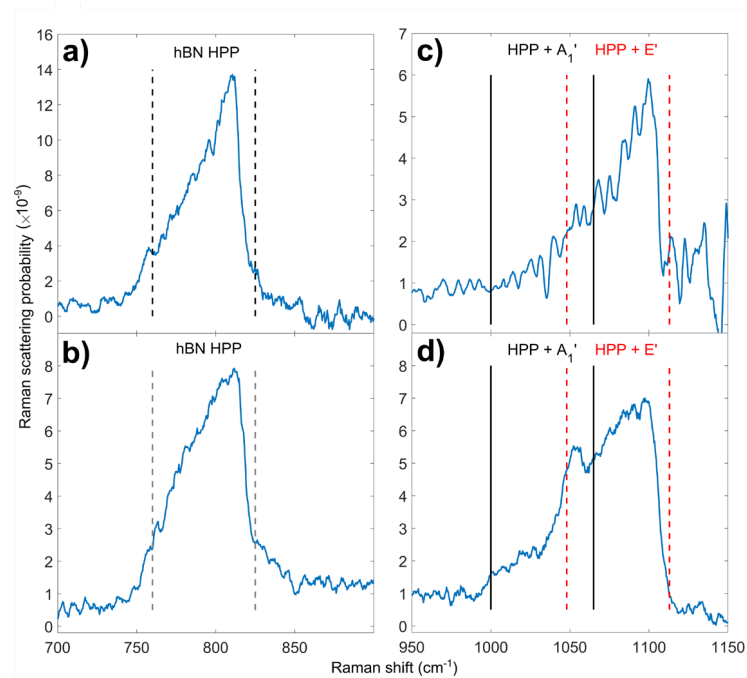

Figure 4: Raman spectra of the hBN encapsulated $MoSe_2$ monolayer sample. Panels a) & b) show the broad 800 $cm^{-1}$ hBN Raman peak with Reststrahlen band shifts marked with dashed lines. Panels c) & d) show the spectra around 1050 $cm^{-1}$, with markings at the Reststrahlen band shifts plus the $MoSe_2$ $A_1'(\Gamma)$ phonon (240 $cm^{-1}$) in black and plus the $MoSe_2$ $E'(\Gamma)$ phonon (288 $cm^{-1}$) in red. Panels a) & c) are from spectra taken around the A1s outgoing / A2s incoming resonances at 1.80 eV. Panels b) & d) are from spectra taken with the laser energy between the B1s outgoing resonance and the B2s incoming resonance at 2.01 eV.

Whilst the attribution of the hBN features to HPP is not new, so far no one has tried to understand the lineshape in terms of this model. In order to fully model the Raman lineshape would require a model not only for the HPPs but also a model for TMD excitons and for the HPP-exciton interaction.

However, it is reasonable to expect that the density of states of the HPP may well dominate the lineshape. Therefore we have used a T-matrix model, developed by Passler et al.[47], to predict the IR absorption spectrum of HPP for our structure, measured in an Otto configuration experiment, as a function of in-plane wavevector of the absorbed light. The absorption spectrum predictions were made using the hBN dielectric function determined by Caldwell et al.[16] and layer thicknesses constrained by atomic force microscopy and fitting to the optical reflectivity spectra. The predicted spectra were fitted to extract the dispersion relation for the HPP, i.e. the energy of the HPP as a function of in-plane wavevector. For some in-plane wavevectors it is possible to observe multiple, different energy, HPP which have different mode profiles in the direction normal to the plane. As shown in Fig 5a as the in-plane wavevector increases the number of possible modes increases and the energy of a specific mode decreases. In a Raman scattering momentum conservation should apply to the wavevector of the incoming photon and the sum of the wavevectors of the outgoing phonon and HPPs generated during the scattering process. This places a limit on the maximum wavevector HPP that can be generated by Raman scattering comparable to the wavenumber of visible light ~15000 cm$^{-1}$. If we use the wavelength of light resonant with the A2s exciton in WSe$_2$ and based upon back scattering of radiation incident at 30 °, reasonable for our objective lens, we predict that the maximum wavevector HPP we should generate in momentum conserving Raman scattering would have a magnitude of 15050 cm$^{-1}$. As shown on Fig 5 HPP with wavevectors with magnitude are restricted to the energy range 800-820 cm$^{-1}$ which is clearly too small a range to explain the observed Raman feature which extends below 750cm$^{-1}$. Defects and other effects can lead to non-momentum conserving Raman scattering. As shown on Fig 4, if we assume the maximum wavevector of HPP observed in Raman scattering is five times that of the momentum conserving limit the range of the HPP Raman feature would be predicted to increase significantly to 760-820 cm$^{-1}$. Therefore, it is difficult to believe that Raman scattering does not probe HPP with wavevectors significantly bigger than the momentum conserving limit of 15050 cm$^{-1}$. However even if we place no upper limit on the wavevectors of the HPP which can be detected by Raman the range the theory predicts a feature which is narrower than observed. Fundamentally, it is not possible for the current model to produce scattering below the lower Reststrahlen band edge. Thus, whilst it is highly likely that HPP are responsible for the two features associated with them in Raman spectra of hBN encapsulated TMDs layers the simple HPP model set out above cannot explain the width of these features.

We now need to consider why the simple HPP model fails and what additional physics is required to explain the observed Raman features. An obvious possibility is that we have used the incorrect parameters for the Reststrahlen band. However a review of the literature gives multiple independent measurements of these parameters based on IR spectroscopy[16,37,48,49] and whist there are variations in the extracted parameters none of the measurements support the lower bound of the Reststrahlen band being below 760 cm$^{-1}$. Another alternative is that the 730-850 cm$^{-1}$ feature is a combination peak involving absorption of a thermal phonon. However, the measurements were performed at 4 K and the thermal energy, which corresponds to ~3 cm$^{-1}$, is insufficient to significantly extend the feature to lower shifts. Another possibility is that at very large wavevectors the ZO phonon responsible for the HPP disperses to lower energy. However, there is no obvious mechanism allowing us to couple to such large wavevector HPP. Thus, currently we do not have a good explanation of the width of the two Raman features associated with HPP in these structures. Whilst it is possible, the resolution of this issue is associated with the TMD layers it is also possible that the ability of Raman scattering to access large wavevector HPP is allowing us to access new HPP effects not yet understood.

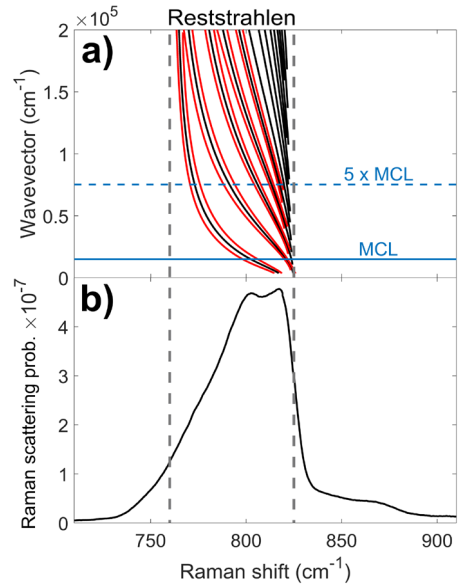

*Figure 5: a) Dispersion relation for the hyperbolic phonon polariton modes, determined using the 4x4 transfer matrix approach of Passler et. al.[47] The black lines show the bands of allowed modes and the red lines indicate their widths (at half magnitude) for the first five allowed HPP modes. The maximum in-pane momentum available in a single phonon Raman process based on the incident angle of the beam from the 0.5 NA objective is 15050 cm$^{-1}$ at the A2s energy. This momentum conserving limit is marked with a solid blue line labelled MCL. Five times this limit is also marked with a dashed blue line. b) shows a Raman spectrum taken on the hBN encapsulated WSe$_2$ monolayer sample with an excitation energy of 1.866 eV, on resonance with the WSe$_2$ A2s exciton. The hBN TO and LO frequencies of 760 and 825 cm$^{-1}$ are marked with dashed vertical lines, between which is the Reststrahlen band.*

The results set out in this paper add considerably to our knowledge about the hBN HPP Raman features observed previously in hBN encapsulated WSe$_2$ structures. In WSe$_2$, we have observed outgoing resonances with the A2s and A3s excitonic states for the first time. The outgoing resonance with the A3s exciton is the first time this exciton has been detected in resonance Raman. In MoSe$_2$, we have observed hBN HPP features for the first time. In this material, they are associated with double resonances with the incoming photons resonant with the A2s or B2s excitons and the outgoing photons resonant with the A1s or B1s excitons respectively. From the MoSe$_2$ spectra, we can deduce that in this material the upper HPP feature involves emission of a HPP and either an $A_1'$ or $E'$ phonon, with both cases contributing to the overall lineshape. It is quite likely that this is also true for WSe$_2$ but the degeneracy of the phonons in this material makes this difficult to confirm.

However, the most important outcomes of this study are associated with the HPP. Firstly, it is clear that compared to IR reflectivity measurements, resonance Raman scattering allows access to a much greater range of HPP with larger wavevectors. Even momentum conserving Raman allows access to wavevectors a factor of ten greater than accessible by IR measurements and there is good evidence that Raman scattering is not momentum conserving allowing even larger wavevector modes to be studied. In addition, there is a mystery as to how Raman scattering can occur at Raman shifts below the lower bound on the Reststrahlen band determined from IR measurements. The solution to this mystery may give us insights into HPP that cannot be achieved via other techniques. The insights need not be restricted to HPP associated with the layers of hBN. It is possible to transfer exfoliated TMD flakes onto a wide range of substrates. This suggests they could make an ideal probe for HPP associated with a wide range of materials and metamaterials. This paper shows that doing this will give an insight into HPP in these structures that is difficult to achieve in other ways.

## Methods

The hBN encapsulated monolayer MoSe$_2$ and WSe$_2$ samples were fabricated by hot pickup using a dry polycarbonate on PDMS stamp from individual mechanically exfoliated monolayers[50]. The WSe$_2$ sample was encased within a top 36 nm, and bottom 27 nm, thick hBN layer. The bottom hBN layer was on top of a 20 nm graphite layer which sat atop a 300 nm thick SiO$_2$ coated Si wafer. The MoSe$_2$ sample followed the same structure, with top and bottom hBN thicknesses of 30 and 20 nm respectively and a 4 nm graphite thickness. These are the same samples that feature in a larger resonance Raman study[43,51] with the same experimental setup. The measurements presented here were performed with the sample under vacuum at 4 K in a helium flow cryostat. A Coherent Mira 900 Ti:Sapphire laser in CW mode and a Coherent CR-599 CW dye laser were used to excite the samples in backscattering geometry via a 50x 0.5 NA long working distance microscope objective. This allowed Raman spectra to be taken with excitation energies from 1.60 to 2.24 eV. The incident power on the sample was 100 µW. The Raman spectra were measured with a TriVista 555 spectrometer with the first two stages of the spectrometer set in subtractive mode and a liquid nitrogen cooled CCD.

Linear polarisers were placed in the exciting laser beam before the sample and in between the sample and spectrometer. For each excitation energy, Raman spectra were taken with these polarisers set co-linear and with them crossed. The crossed polariser spectra were subtracted from the co-linear spectra which had the effect of removing the photoluminescence (PL) signal whilst preserving the Raman signal. At energies where PL signal in the spectrum was many orders of magnitude greater than the Raman, such as near the A1s exciton energy, residual PL was present after subtraction.

The Raman signal from the SiO$_2$ coated Si substrate was used to calibrate the Raman spectra for frequency. The intensity of the Si peak was used with Si resonance Raman data from literature[52] to convert the data to absolute scattering probability. A transfer matrix model of the stack of layers in the samples was used to correct for the effect of thin film interference on the Raman intensity. The details of these corrections are given in the supplementary information of our previous paper[43].

## Supplementary Information

The supplementary information[39] contains discussion of the hBN modes in a hBN encapsulated MoSe$_2$ monolayer sample, including a resonance Raman colourmap and resonance profile of the 800 cm$^{-1}$ mode. Reflectivity spectra are also discussed for the MoSe$_2$ and WSe$_2$ samples, with exciton energies extracted from those spectra presented with comparative literature values from[44,53–56]. Additional detail on the extraction and fitting of resonance profiles from the Raman spectra is also provided.

## Data Availability

The data presented in this paper is openly available from the University of Southampton Repository DOI: https://doi.org/10.5258/SOTON/D1734

## Corresponding Author

*D.C.Smith@soton.ac.uk

**Author Contributions**

Samples were fabricated by P.R. The experimental measurements were performed by J.V and L.P.M. Experimental data analysis and interpretation was carried out by J.V., L.P.M and D.C.S. The paper was written by D.C.S and J.V. All authors discussed the results and commented on the manuscript.

The manuscript was written through contributions of all authors. All authors have given approval to the final version of the manuscript.

**Funding Sources**

Research at the University of Southampton was supported by the Engineering and Physical Science Council of the UK via programme grant EP/N035437/1. Both L.P.M and J.V were also supported by EPSRC DTP funding. The work at University of Washington was mainly supported by the Department of Energy, Basic Energy Sciences, Materials Sciences and Engineering Division (DE-SC0018171).

# Supplementary information for Insights into hyperbolic phonon polaritons in hBN using Raman scattering from encapsulated transition metal dichalcogenide layers

Jacob J.S. Viner [1], Liam P. McDonnell [1], Pasqual Rivera [2], Xiaodong Xu [2], David C. Smith [1*]

[1] School of Physics and Astronomy, University of Southampton, Southampton SO17 1BJ, United Kingdom.

[2] Department of Physics, University of Washington, Seattle, WA, USA

**Table of Contents**

## S1) Details Resonance Model Fitting Procedure

The resonance profiles of these broad hBN peaks were obtained by averaging the intensity over a range of 5 cm$^{-1}$ centred at a chosen Raman shift for each Raman spectrum. It is this averaged Raman scattering probability, $I(E)$, as a function of excitation energy, that was fitted to the 3 state third order perturbation theory model for absolute scattering probability:

$$I(E) = \left| \sum_{i,j=1}^{3} \frac{a_{ij}}{(E - E_i - i\Gamma_i)(E - E_j - E_{p1} - i\Gamma_j)} \right|^2 \quad (1)$$

Where $E_i$ is the energy of the state $i$, and $\Gamma_i$ is the width; $E_{p1}$ is the phonon energy; $a_{ij}$ represents the matrix element for the scattering channel involving states $i$ and $j$; and $E$ is the excitation energy. This amplitude $a_{ij}$ is complex, and during the fitting process it is treated as a real amplitude multiplied by $e^{i\theta}$ where $\theta$ is between 0 and 2π. The resonance profiles at the chosen centre Raman shifts of 750, 775, 815, 1000, 1025 and 1050 cm$^{-1}$, represented in Figure S1 by the coloured bands, were fitted separately.

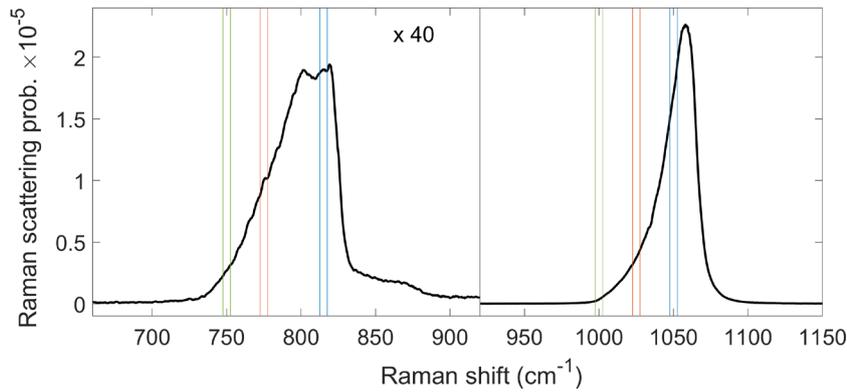

*Figure S1: Subsection of the Raman spectrum presented in Fig 1 of the main body of the paper. The spectrum was taken with an excitation energy of 1.866 eV on the hBN encapsulated monolayer WSe$_2$ sample. The data below 930 cm$^{-1}$ was scaled by a factor of 40 for comparison. The bands of energies averaged over to extract the resonance profiles presented in the paper are marked with pairs of coloured lines.*

The best-fit parameters for the six WSe$_2$ resonance profiles presented in Fig 2 in the main body of the paper are presented in Table S1. Here, the real part of the scattering matrix element squared corresponds to the Raman scattering probability. The energies of the three excitons from the fits to all six resonance profiles all agree to within 6 meV. The widths are generally in good agreement; to within two standard errors of each other. Additional details on the derivation of the resonance model are given in the supplementary information of our previous work[S1].

|  | Scattering Channel | Raman shift (cm$^{-1}$) | | | | | |
|---|---|---|---|---|---|---|---|
|  |  | 750 | 775 | 815 | 1000 | 1025 | 1050 |
| **Absolute Raman scattering probability (x10$^{-16}$)** | A1s - A1s | 2.29±0.95 | 7.4±1.2 | 32±3 | 189.5±8.4 | 60.6±43 | 81±2400 |
|  | A1s - A2s | 9.55±0.02 | 40.5±0.12 | 70.0±0.14 | 17.65±0.16 | 56.2±0.3 | 986±13 |
|  | A1s - A3s | 3.10±0.36 | 50.4±0.67 | 69±1 | 7.3±1.1 | 300±17 | 100±960 |
|  | A2s - A2s | 0.062±3.9 | 0.13±23 | 97±2.7 | 409±83 | 350±780 | 1000±66000 |
|  | A2s - A3s | 16.6±1.3 | 13±19 | 19.1±2.0 | 71.6±45 | 1820±370 | 0±22000 |
|  | A3s - A3s | 6.1±2.7 | 15±34 | 99±7.1 | 756±21 | 560±250 | 1300±3400 |
| **Energy (eV)** | A1s | 1.736±0.001 | 1.736±0.005 | 1.738±0.001 | 1.734±0.001 | 1.732±0.001 | 1.736±0.001 |
|  | A2s | 1.868±0.0004 | 1.867±0.0004 | 1.867±0.0002 | 1.869±0.001 | 1.869±0.001 | 1.865±0.003 |
|  | A3s | 1.891±0.003 | 1.890±0.007 | 1.892±0.001 | 1.895±0.017 | 1.894±0.003 | 1.897±0.005 |
| **Width (meV)** | A1s | 6.0±1.1 | 6.4±0.4 | 6.5±0.8 | 8.7±1.1 | 5.9±0.3 | 7.9±1.1 |
|  | A2s | 6.0±0.5 | 5.6±0.3 | 5.3±0.2 | 7.0±0.7 | 6.9±0.4 | 7.0±0.1 |
|  | A3s | 10.0±7.0 | 10.8±1.1 | 10.5±1.4 | 8±14 | 4.8±2.0 | 4.2±2.0 |

*Table S1: Fitting parameters from the resonance Raman profiles taken at different Raman shifts across the 800 and 1050 cm$^{-1}$ broad hBN Raman features in the hBN encapsulated monolayer WSe$_2$ sample. The corresponding profiles are shown in Figure 2 in the main body of the paper. The errors given are standard deviations determined from the fitting process.*

**S2) Monolayer MoSe$_2$ Resonance Raman**

Figure S2 shows a colourmap of resonance Raman data on an encapsulated MoSe$_2$ monolayer. In this material Raman peaks due to single gamma-point phonons were observed at 240 cm$^{-1}$ for the A$_1$'(Γ) and 290 cm$^{-1}$ for the E'(Γ). Numerous further Raman peaks were observed for this material with frequencies up to 600 cm$^{-1}$. Comparing the frequencies of the observed peaks with published phonon dispersion relations[S2], peaks above 350 cm$^{-1}$ must necessarily be multiphonon peaks. A list of possible assignments to all of the peaks observed is given in the supplementary information of our previous report on WSe$_2$ and MoSe$_2$ monolayers[S1]. No Raman peaks above 600 cm$^{-1}$ have been reported in literature for unencapsulated monolayer MoSe$_2$ samples[S3–5], so there should be no scattering due to multiphonon Raman at the shifts covered by the 800 and 1050 cm$^{-1}$ hBN modes.

For this MoSe$_2$ sample, the scattering from the TMD associated Raman peaks, which fall at Raman shifts of 600 cm$^{-1}$ or less, was generally more intense than that from the hBN modes. The TMD Raman peaks show resonant enhancement at the A1s, A2s, B1s and B2s excitons as described in our previous work[SS1], with the 2s states producing weaker resonances due to their lower oscillator strength. In Figure S2a) the incoming and outgoing resonances associated with each exciton are indicated by the solid and dashed lines respectively with the colour of the lines indicated if the state is an A (white) or B (red) excitonic state. The incoming resonances at the energies of the excitons are marked with solid lines and the outgoing resonances, at the exciton energy plus the phonon energy, are indicated with dashed lines. The Si Raman peak from the sample substrate is present at 520 cm$^{-1}$ across the entire range of excitation energies used.

Scattering from the broad 800 and 1050 cm$^{-1}$ hBN modes is observed around 1.8 eV, falling between the A1s outgoing and A2s incoming resonances which are indicated by white lines. Example Raman spectra showing the lineshape of the modes at 1.8 eV are given in Figure S2c) & e). Strong PL signal which follows the A1s outgoing resonance appears with the feature at 1050 cm$^{-1}$ between the A1s outgoing and A2s incoming resonance energies. Scatter from these hBN modes is also visible around 2.0 eV which falls between the B1s outgoing and B2s incoming resonances, indicated in Red. The lineshapes at this energy are shown in Figure S2b) & d). Notably, no clear resonance is observed in the colourmap between the B1s incoming resonance and the A2s outgoing resonance. Unlike the WSe$_2$ case (see Fig 3 in main paper) in the monolayer MoSe$_2$ spectra we do not observe any features near 750 cm$^{-1}$ which could be associated with multi-phonon TMD Raman modes. In MoSe$_2$ we also note the intensity of the hBN peaks is strongest at double resonances associated with a 1s outgoing and 2s incoming resonance, with minimal scattering observed between A and B excitons. It is also interesting to note that the 1050 cm$^{-1}$ feature is less intense than the one at 800 cm$^{-1}$, unlike the WSe$_2$ case. This is despite being much closer to a double resonance condition than the 800 cm$^{-1}$ feature.

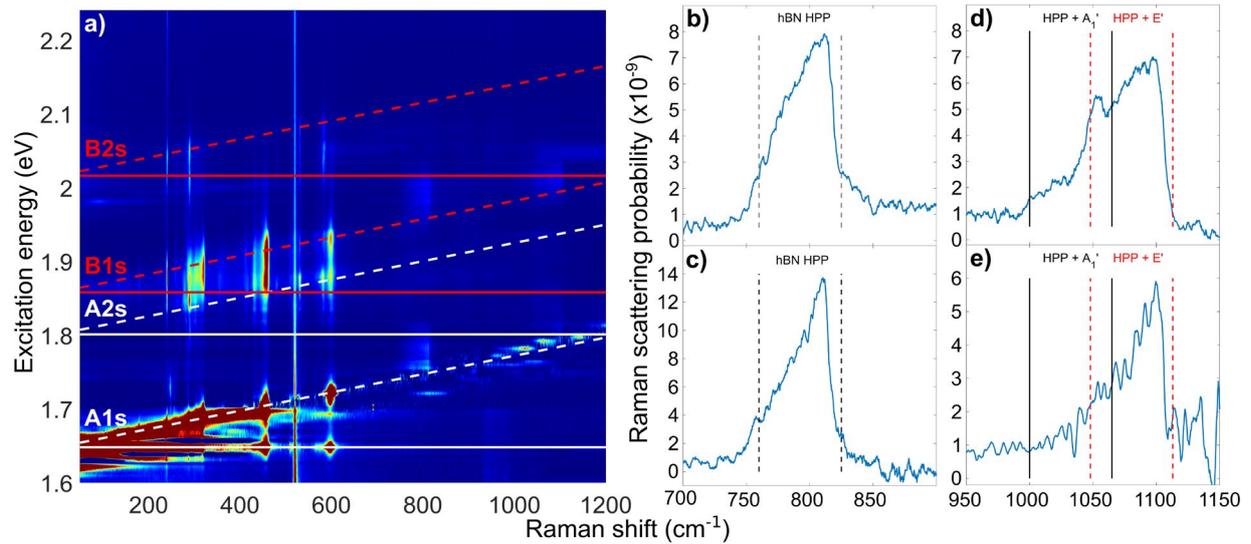

*Figure S2: a) Colourmap of a resonance Raman spectra taken at 4 K of monolayer MoSe$_2$ encapsulated in hBN with excitation energies from 1.6 to 2.25 eV. The resonance energies associated with the A1s and A2s excitons are shown in white and the B1s and B2s shown in red, with solid lines corresponding to the incoming resonance and the dashed lines corresponding to the outgoing resonances at the exciton energy plus the phonon energy. The intensity of the Raman scattering is indicated in logarithmic scale by the colour, with dark red corresponding to most intense and dark blue least intense. b)-e) subsections of Raman spectra of MoSe$_2$ reproduced from the main body of the paper showing the lineshape of the two hBN HPP associated modes. The upper two plots correspond to an excitation energy of 2.01 eV and the lower two plots are 1.80 eV.*

In Figure S3a resonance Raman profile for the 800 cm$^{-1}$ hBN peak in MoSe$_2$ is shown. This profile was extracted, by averaging the intensity of the Raman signal for shifts between 797.5 and 802.5 cm$^{-1}$, for each of the spectra. The strongest signal is present between the 1s outgoing and 2s incoming resonances. This is true for both the A and B excitons and suggests strong inter-state scattering for A1s-A2s and B1s-B2s. The signal at the 1s incoming and 2s outgoing energies is not as intense, suggesting that the amplitudes of the scattering involving a single excitonic state are small.

The data was fitted to a four-state resonance model. The best fit obtained is shown on the figure. The four-state model takes the same form as the three-state model used for WSe$_2$, given in equation (1), with the sum taken up to 4 rather than 3. This model requires 10 scattering terms, one for scattering via each of the 4 excitons individually, and 6 corresponding to inter-state scattering.

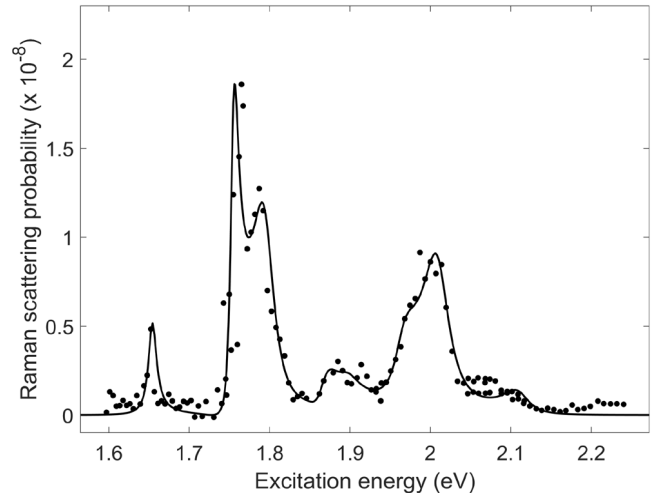

*Figure S3: Resonance Raman profile extracted for the MoSe$_2$ hBN 800 cm$^{-1}$ peak shown in Figure S2 with a four-state resonance Raman scattering model fitted to the data. The excitation energies used range from 1.60 – 2.24 eV, spanning from below the MoSe$_2$ A1s exciton to above the B2s exciton.*

Whilst the fitted profile shows reasonable qualitative agreement with the resonance data the uncertainties associated with the fitted amplitudes are orders of magnitude larger than the obtained amplitude coefficients. This arises due to overfitting of the resonance data, where the large number of amplitude terms required in the four-state model can take multiple possible values, preventing the fit from converging towards a single solution. This is partly a consequence of the fact that the amplitude terms can combine in both addition and subtraction, depending on their phase, leading to multiple possible solutions. The parameter correlation matrix for the fitted model is illustrated in Figure S4 as a colour-map. The top left hand 10x10 square of high intensity corresponds to the correlation between the amplitude terms in the fit. The 10x9 area at the top right corresponds to the correlation between the amplitude and phase terms and the 9x9 area in the bottom right corner is the correlation between the phase terms. The remaining area corresponds to correlation with the fitted energy and width terms with themselves and all other parameters. The 'blue cross' indicates that there are no strong correlations between the energies and widths and all the other parameters. Thus, the uncertainties in amplitude and phase do not affect the ability of the fit to constrain the energies and widths. This is also reflected in the confidence intervals for these parameters that are of the order of 10 meV.

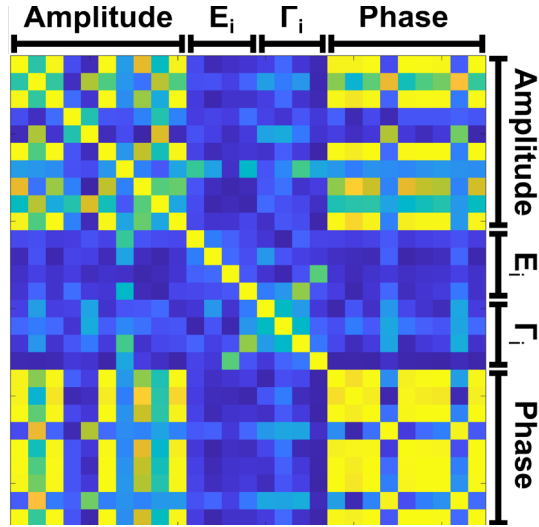

*Figure S4: Visualisation of the magnitude of the parameter correlation matrix for the fitting parameters in the 4-state resonance fit. Blue is close to zero (no correlation) and yellow is close to 1 (maximum correlation). The first 10 terms are the amplitudes, the next four are the energies of the excitonic states followed by the four widths. The final 9 are the phases of the scattering channels, where the A2s-A2s channel is set to zero phase.*

These fitted energies of the excitonic states all agree to within 6 meV of the values determined from fitting resonance Raman profiles of the 240 cm$^{-1}$ $A_1'(\Gamma)$ MoSe$_2$ phonon. For this phonon, the resonance data was fitted to a single state resonance model at the A1s, a two-state model at the A2s and B1s and another single state model at the B2s[SS1]. The energies and widths are given for both of these cases in Table S2. The widths are generally broader for the hBN HPP fit however, the much larger uncertainties mean that it is not significant.

| Exciton | MoSe$_2$ fitted exciton energy (eV) | | MoSe$_2$ fitted exciton width (meV) | |
|---|---|---|---|---|
| | $A_1'(\Gamma)$ phonon profiles | hBN HPP profile | $A_1'(\Gamma)$ phonon profiles | hBN HPP profile |
| A1s | 1.648 ± 0.001 | 1.654 ± 0.003 | 3.0 ± 0.1 | 6.2 ± 3.3 |
| A2s | 1.804 ± 0.001 | 1.796 ± 0.013 | 8.2 ± 1.7 | 17.0 ± 7.5 |
| B1s | 1.858 ± 0.001 | 1.860 ± 0.020 | 18.9 ± 1.3 | 17.1 ± 15.2 |
| B2s | 2.016 ± 0.001 | 2.012 ± 0.010 | 9.4 ± 0.7 | 21.5 ± 14.0 |

*Table S2: Energy and width coefficients of the A1s, A2s, B1s and B2s excitons from fitting the broad 800 cm$^{-1}$ hBN mode resonance Raman profile for the hBN encapsulated MoSe$_2$ monolayer. These are given alongside equivalent energies and widths determined by fitting the resonance Raman profiles of the 240 cm$^{-1}$ $A_1'(\Gamma)$ MoSe$_2$ phonon from the same Raman spectra. The errors given are standard deviations determined from the fitting process.*

## S3) Monolayer MoSe$_2$ Reflectivity

A reflectivity spectrum of the MoSe$_2$ monolayer was taken under the same conditions as the Raman spectra. The reflectivity was normalised to the silicon substrate and then the spectrum was differentiated to enhance just the contributions to the reflected signal coming from the monolayer. The differential reflectivity spectrum is shown in Figure S5. Three features were observed in the which were assigned to the A1s, A2s and B1s excitons. The B2s was not clearly observed which is most likely as a result of the broader linewidth observed at the B excitons combined with and lower oscillator strength of the 2s states.

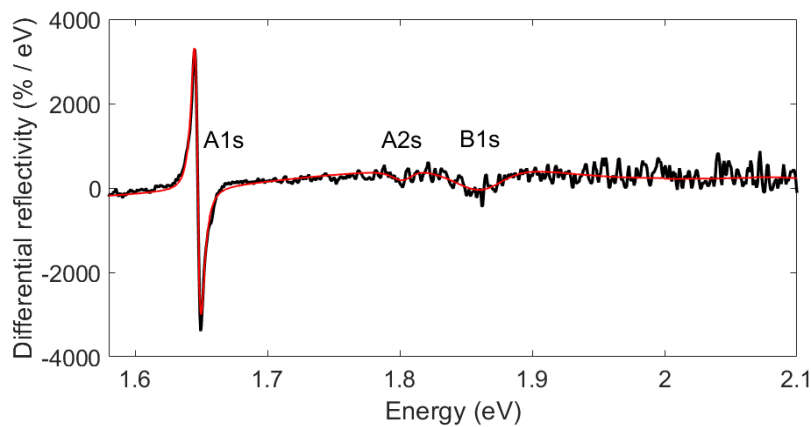

*Figure S5: Differential reflectivity spectrum of the MoSe$_2$ sample relative to the Si Substrate. The best fit t-matrix model for the spectrum is shown in red.*

A transfer matrix model was used to model sample, with the permittivity of the MoSe$_2$ layer comprised of 3 Lorentzian oscillators. Best fit values were found for the amplitude, linewidth and energy of these oscillators by fitting the reflectivity spectrum. The red curve on Figure S5 shows this fitted model which shows good agreement with the spectrum. The best fit values are given in Table S3. The errors shown are the standard deviations determined from the fitting process. All three of the states observed in the reflectivity fell within one standard deviation of the value determined from Raman scattering. Further, the A1s, B1s and B2s energies fell within the range of energies reported for these states in literature.

|         | Energy (eV)       |                   |                  |
|---------|-------------------|-------------------|------------------|
| Exciton | Raman             | Reflectivity      | Literature[S6,7] |
| A1s     | 1.648 ± 0.001     | 1.648 ± 0.001     | 1.642, 1.655     |
| A2s     | 1.803 ± 0.002     | 1.804 ± 0.003     | 1.793, ---       |
| B1s     | 1.858 ± 0.001     | 1.856 ± 0.003     | 1.839, 1.866     |
| B2s     | 2.016 ± 0.001     | ---               | 2.000, 2.019     |

*Table S3: Comparison of fitted MoSe$_2$ exciton energies from the resonance Raman and reflectivity measurements as well as published values in literature.*

## S4) Monolayer WSe₂ Reflectivity

Figure S6 shows the reflectivity spectrum of the WSe$_2$ monolayer sample. Features are present at the A1s, A2s and A3s exciton energies. Whilst those features at the A1s and A2s are obviously present in the data, the lower oscillator strength of the A3s means it is less clear. The transfer matrix model was used to fit this spectrum with 3 Lorentz oscillators and the best fit curve is shown in red on the figure.

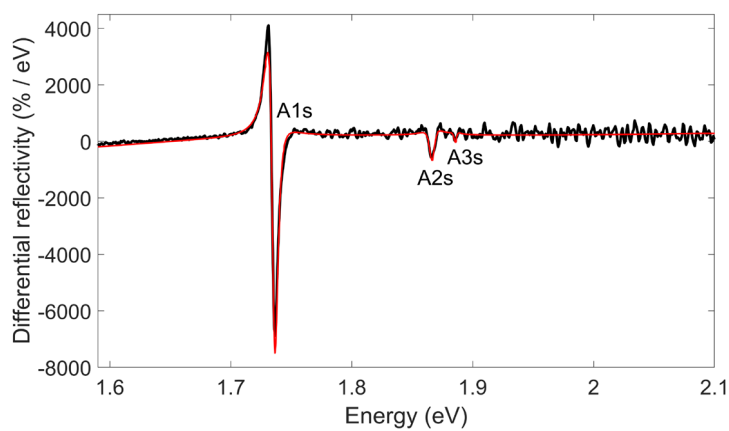

*Figure S6: Differential reflectivity spectrum of the WSe$_2$ sample relative to the Si Substrate. The best fit t-matrix model for the spectrum is shown in red.*

The best fit values from the model are given in Table S4 alongside values obtained from fitting resonance Raman profiles. For the A1s and A2s, the energies obtained from fitting the profile of the singe-phonon peak at 250 cm$^{-1}$ are given. For the A3s, where that 250 cm$^{-1}$ peak was not clearly visible, the energy obtained from the resonance profile for the hBN HPP is given. The energies from reflectivity are all 3-4 meV below those from the resonance Raman. However, the spacing of the states in the Raman and reflectivity cases are the same to within 1 meV or better. Comparing the energies to values from literature reveals that the energies from this sample are on the higher end of those reported previously. The much more consistent energy spacing between the states of 131 meV and 20 meV for the A1s-A2s and A2s-A3s respectively are within 1 meV of previously reported values 131-132 meV and 21-22 meV which strongly suggests our assignment of the excitons is correct.

|         | Energy (eV)     |                 |                      |
|---------|-----------------|-----------------|----------------------|
| Exciton | Raman           | Reflectivity    | Literature[S8–10]    |
| A1s     | 1.740 ± 0.001   | 1.737 ± 0.001   | 1.723, 1.712, 1.741  |
| A2s     | 1.871 ± 0.001   | 1.867 ± 0.001   | 1.855, 1.843, 1.871  |
| A3s     | 1.891 ± 0.002   | 1.887 ± 0.001   | 1.877, 1.864, ---    |

*Table S4: Comparison of fitted WSe$_2$ exciton energies from the resonance Raman and reflectivity measurements as well as published values in literature.*